\documentclass[twocolumn]{aastex631}

\usepackage{amsmath}
\usepackage{amssymb}
\usepackage{graphicx}
\usepackage{natbib}
\usepackage{subfigure}   
\usepackage{color}
\usepackage{tabularx}
\usepackage{rotating}
\usepackage{multirow}
\usepackage{ulem}
\usepackage{float}
\usepackage{url}
\usepackage[utf8]{inputenc}

\begin{document}

\title{Classification for 969 double-mode RR Lyrae stars from Zwicky Transient Facility}
\author[0009-0001-5015-0387]{Jianxing Zhang}
\affiliation{CAS Key Laboratory of Optical Astronomy, National Astronomical Observatories, Chinese Academy of Sciences, Beijing 100101, China}
\affiliation{School of Astronomy and Space Science, University of the Chinese Academy of Sciences, Beijing, 100049, China}

\author[0000-0001-7084-0484]{Xiaodian Chen}
\affiliation{CAS Key Laboratory of Optical Astronomy, National Astronomical Observatories, Chinese Academy of Sciences, Beijing 100101, China}
\affiliation{School of Astronomy and Space Science, University of the Chinese Academy of Sciences, Beijing, 100049, China}
\affiliation{Institute for Frontiers in Astronomy and Astrophysics, Beijing Normal University, Beijing 102206, China}

\author[0000-0003-4489-9794]{Shu Wang}
\affiliation{CAS Key Laboratory of Optical Astronomy, National Astronomical Observatories, Chinese Academy of Sciences, Beijing 100101, China}
\affiliation{School of Astronomy and Space Science, University of the Chinese Academy of Sciences, Beijing, 100049, China}

\author[0009-0002-1155-284X]{Jiyu Wang}
\affiliation{CAS Key Laboratory of Optical Astronomy, National Astronomical Observatories, Chinese Academy of Sciences, Beijing 100101, China}
\affiliation{School of Astronomy and Space Science, University of the Chinese Academy of Sciences, Beijing, 100049, China}

\author[0000-0001-9073-9914]{Licai Deng}
\affiliation{CAS Key Laboratory of Optical Astronomy, National Astronomical Observatories, Chinese Academy of Sciences, Beijing 100101, China}
\affiliation{School of Astronomy and Space Science, University of the Chinese Academy of Sciences, Beijing, 100049, China}

\correspondingauthor{Xiaodian Chen}
\email{chenxiaodian@nao.cas.cn}

\begin{abstract}
RR Lyrae (RRL) variable stars are cornerstone distance indicators. In particular, double-mode RR Lyrae (RRd) stars enable period–luminosity relations (PLRs) that are less sensitive to metallicity, reducing systematic biases in distance measurements. However, their utility has been limited by a global sample of only $\sim$3,000 objects. We develop an automated RRd-screening pipeline and apply it to a cross-matched sample between the Gaia DR3 RRL catalog and ZTF DR22 time-series photometry. The workflow combines Lomb--Scargle period searches, iterative pre-whitening, period-ratio constraints that suppress $\sim$1-day sampling aliases, and amplitude-based quality cuts, enabling large-scale RRd star screening. We produce two ZTF-based catalogs: (i) 39,322 reliable single-mode RRL (40.5\% of the cross-matched set) and (ii) 969 RRd stars. Among the RRd stars, 614 objects are newly identified, substantially enlarging this previously scarce sample; the catalog achieves an estimated completeness of 47.7\%. The PLR derived from the newly discovered RRd stars agrees with the LMC-based relation, though with larger uncertainties. Incorporating these stars will help tighten the RRd PLR and improve distance measurements. Looking ahead, systematic RRd searches with upcoming surveys such as the Legacy Survey of Space and Time (LSST) and the China Space Station Telescope (CSST) should further extend high-accuracy distances across the Local Group and strengthen their cosmological applications.

\end{abstract}

\section{Introduction}
\label{section1}
\par
RR Lyrae (RRL) variable stars, as a class of old core-helium-burning horizontal branch pulsating variable stars \citep{2020A&A...641A..96S}, are important standard candles in astrophysics \citep{2016ApJ...832..210B,2020JApA...41...23B}. They are widely used to measure distances to nearby galaxies \citep{2022Univ....8..191M,2017AcA....67....1J,2012AJ....144..106H}, characterize the ancient structure of the Galactic halo and streams \citep{2022ApJS..258...20A,2022MNRAS.513.1958W}, and constrain the Hubble constant \citep{2016ApJ...832..210B} due to their accurate empirical period-luminosity relation (PLR) \citep{1986MNRAS.220..279L,2001MNRAS.326.1183B,2015ApJ...807..127M,2023ApJ...945...83M}. Classical RRL are mainly divided into fundamental-mode pulsating RRab stars and first-overtone-mode pulsating RRc stars. Additionally, there exists a class of double-mode variable stars exhibiting both fundamental and first-overtone mode pulsations, known as RRd stars \citep{2001AJ....122..207B,2015ApJS..219...25J,2000AcA....50..491P}. Although RRd stars constitute a small fraction (approximately 1-10\%) of RRL \citep{2019AcA....69..321S,2023A&A...674A..18C}, their double-mode pulsation characteristics endow them with special value.

Traditional RRL distance measurement methods primarily rely on the main component, RRab stars. However, the PLR of RRab stars is significantly affected by metallicity ([Fe/H]) and has considerable intrinsic scatter, limiting their distance measurement accuracy \citep{2023ApJ...945...83M}. Recent studies have found that RRd stars demonstrate significant potential advantages: a linear relationship exists between their period ratio (the ratio of the first-overtone period $P_{\rm 1O}$ to the fundamental period $P_{\rm F}$) and metallicity \citep{2015ApJ...808...50M,2023NatAs...7.1081C}, leading to a metallicity-insensitive PLR. If this metallicity-independent PLR can be precisely calibrated and applied, it is expected to significantly reduce errors in distance measurements, providing a more effective tool for detailed studies of the Galactic structure and the cosmological distance scale.

However, the use of RRd stars for high-precision distance measurements is constrained by a fundamental bottleneck: samples with accurately measured periods remain small and unevenly distributed on the sky. In contrast to the $\sim300,000$ sample of Gaia RRL variable stars, only $\sim2000$ RRd stars are known. Existing RRd samples have largely been assembled either through spectroscopic follow-up that was not tailored to this purpose—hence incomplete and of heterogeneous quality—or through labor-intensive manual vetting of light curves from large time-domain surveys (e.g., OGLE \citep{2019MNRAS.487.5584N}, Gaia \citep{2021A&A...653A..61K}, ZTF \citep{2020ApJS..249...18C,2025ApJS..278....2H}). Leveraging space-based precision photometry, \citet{2024MNRAS.529..296N} identified 72 RRd stars with Kepler and refined several Gaia classifications; similar refinements were reported by \citet{2024AJ....168...43V}. Consequently, there is a pressing need for reliable and efficient methods to mine vast catalogs of nominally single-period RRL photometry for secondary-period signals characteristic of RRd stars.

Based on this, this study aims to: develop an automated pipeline based on time-series photometric data, and use the Gaia Data Release 3 (DR3) catalog \citep{2023A&A...674A..18C} combined with high-cadence photometric data from the Zwicky Transient Facility (ZTF) Data Release 22 (DR22) \citep{2019PASP..131a8003M} to screen for RRd star candidates, construct RRL and RRd catalogs, and finally evaluate the accuracy and completeness of the screening pipeline. The ZTF telescopes were selected due to its longer observational baseline and larger number of epochs. After pre-whitening, the noise level in the power spectrum is significantly reduced, thereby improving the ability to resolve the secondary period. Compared to Gaia, this allows us to identify more RRd stars and investigate possible misclassifications in the Gaia RRd catalog. Furthermore, as a Northern Sky facility, ZTF benefits from extensive spectroscopic follow-up programs (e.g., with the DESI telescope \citep{2025arXiv250314745D} ), facilitating our subsequent research on the relationship between metallicity and RRd periods.

The structure of this paper is as follows: Section \ref{section2} details the data sources and preprocessing methods. Section \ref{section3} introduces the core algorithm flow for automated screening of RRd stars, as well as attempts to screen for multiple periods. Section \ref{section4} presents the screening results, including final sample statistics, period-amplitude diagram and Petersen diagram analysis, and an in-depth discussion comparing with the Gaia DR3 RRd sample. Section \ref{section5} discusses the accuracy and completeness of the screening results and explores the factors affecting both. Section \ref{section6} summarizes the study and outlines directions for future work.

\section{Data}
\label{section2}
This study is based on the cross-match between the 271,779 candidates from the Gaia DR3 RR Lyrae variable star catalog and the ZTF DR22 photometric data. Since ZTF can only observe regions of the sky with declinations above $-28^\circ$, our cross-matching results contain only 112,858 RRL candidates. After obtaining the right ascension and declination from the cross-match, we used the astronomical software package ztfquery to retrieve ZTF time-series photometric data in the $r$ band, obtaining light curves for approximately 97,146 sources. Considering the ground-based observational characteristics and data quality of ZTF, we implemented the following systematic cleaning process: First, for groups of photometric points with consecutive exposure time intervals $\Delta t < 0.001$ d, we performed time aggregation, retaining only one median magnitude point to suppress period overfitting; secondly, we applied ZTF source reliability filtering, keeping sources with the official ZTF variable star quality flag ${\rm catflags} < 30000$; then, we removed 3$\sigma$ outliers from the fitting results through an outlier rejection process (specific fitting details are in Section \ref{section3.2}); finally, we set a data volume threshold, retaining only sources with more than 40 effective photometric points after the above filtering to ensure the reliability of the period analysis.

\section{Method}
\label{section3}
The core pipeline for identifying RRd stars is based on the Lomb-Scargle \citep[LS,][]{1976Ap&SS..39..447L,1982ApJ...263..835S} period analysis method. First, LS analysis is applied to the photometric time series data of each target star to generate a power spectrum. The frequency corresponding to the global peak in the power spectrum is identified as the primary pulsation period. Then, this period is used to fit the original light curve with a Fourier series, and the signal of the first period is removed through pre-whitening (i.e., subtracting the fitted model from the observed data to obtain a residual sequence). Subsequently, we perform LS analysis on the residual sequence to search for the second pulsation period in the power spectrum. Details of the screening process for the second period (such as period ratio constraints, alias exclusion, amplitude filtering, etc.) will be elaborated in the following subsections, combined with the physical characteristics of RRd stars and the features of ZTF observations.

\begin{figure*}
\centering
\includegraphics[width=0.9\textwidth]{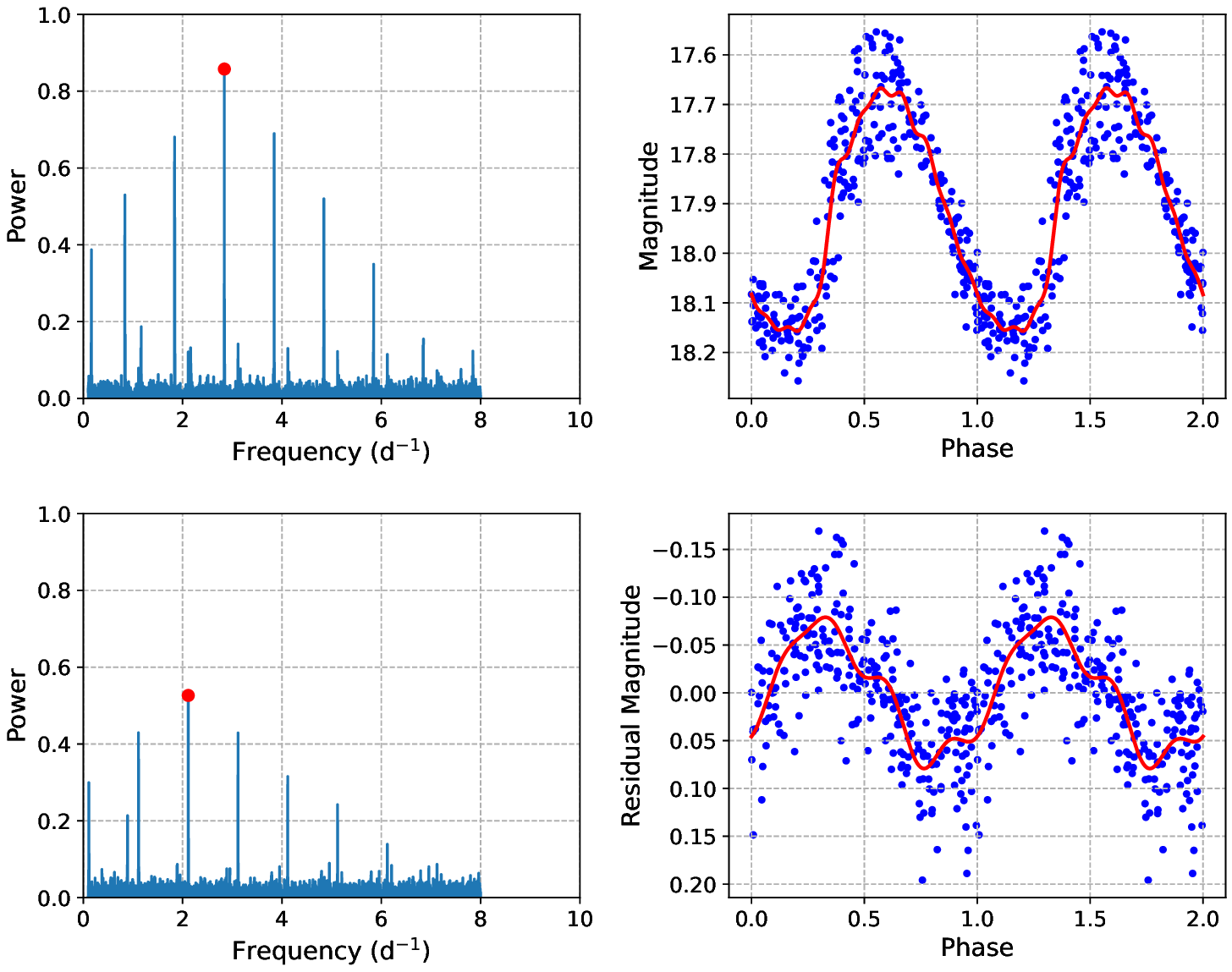}

\caption{This figure shows an example of LS output results. Top left: power spectrum of the photometric data; Top right: light curve folded with the primary period; Bottom left: power spectrum of the residual data; Bottom right: light curve of the residuals folded with the second period.}

\label{Fig.1}
\end{figure*}

\subsection{Period Determination}
\label{section3.1}
After performing LS analysis on all candidates, the globally significant peak in the power spectrum is the primary pulsation period $P_1$. After determining $P_1$, we use a 10th-order Fourier series to fit the original light curve:

\[
f(x) = \frac{A_0}{2} + \sum_{n=1}^{10} \left( A_n \cos \frac{n\pi t}{P} + B_n \sin \frac{n\pi t}{P} \right)
\]
with constraints $0.2 < P_1 < 1$ d, and the amplitude must not be erroneous values of hundreds of magnitudes and must be greater than the photometric error $\sigma$. This step successfully determined reliable $P_1$ for 39,322 RRL, with minimal deviation from the corresponding periods in the Gaia DR3 catalog (typical deviation less than 0.001 d), verifying the period's reliability. Subsequently, we obtained the residual light curve (i.e., the pre-whitening process) by subtracting this 10th-order Fourier fitted model from the observed data, removing the variation signal dominated by $P_1$. LS analysis is performed on the pre-whitened residual sequence to generate a new power spectrum to search for the second period $P_2$. It is known that the period ratio of RRd stars typically follows a specific sequence (see Fig.~\ref{Fig.3}). Therefore, based on the determined $P_1$, we constrained the search for power spectrum peaks within the period ratio interval from 0.72 to 0.75 \citep{2019AcA....69..321S,2023A&A...677A.177N}. If a significant peak with a False Alarm Probability (FAP, the probability that the period is obtained from non-periodic data) less than $10^{-5}$ is detected within this interval, it is considered a credible second period signal.

Crucially, the observed period ratio provides information about the dominant pulsation mode: if $P_2 < P_1$, it indicates that $P_1$ corresponds to the fundamental pulsation mode (i.e., the longer period $P_{\rm F}$), and $P_2$ corresponds to the first-overtone mode ($P_{\rm 1O}$). Conversely, it means $P_1$ is the first-overtone mode, and the found $P_2$ will be its corresponding fundamental mode. This study analyzed and recorded both scenarios. Fig.~\ref{Fig.1} shows the power spectra for the first and second periods, the light curves folded with their respective periods, and their Fourier fitting results for a typical RRd star.

\begin{figure*}
\centering
\includegraphics[width=0.9\textwidth]{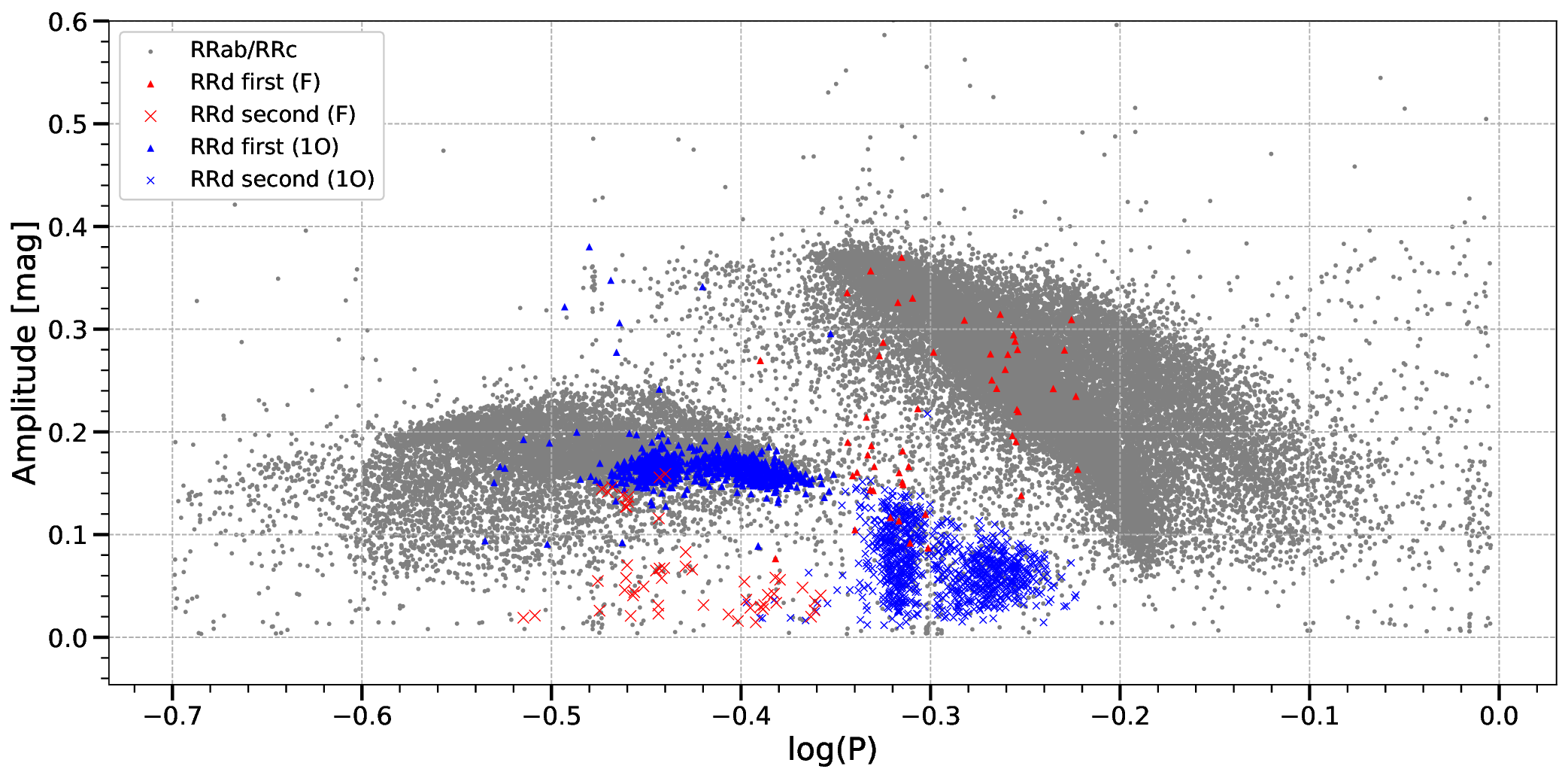}
\caption{Period-amplitude diagram for RRL. Gray points represent RRab and RRc stars. Red points are RRd stars where the primary period is the fundamental mode; blue points are RRd stars where the primary period is the first-overtone mode. Cross symbols mark the corresponding primary period, triangle symbols mark the second period.}
\label{Fig.2}
\end{figure*}

\subsection{Exclusion of False Periods}
\label{section3.2}
After preliminarily identifying double-mode candidates through the steps in Section \ref{section3.1}, a series of strict screenings are required to exclude false signals introduced by the characteristics of ZTF ground-based observations, and constraints are applied based on the physical properties of RRd stars. The first priority is to remove contamination from the daily sampling window and its associated beat frequencies. Because Earth’s rotation imposes an approximately 1-day cadence on ZTF and RRL pulsation periods are themselves close to 1 day, spurious periodogram peaks can easily alias with true signals. Therefore, we excluded candidates meeting any of the following conditions: (1) its first or second pulsation frequency $f_1$ or $f_2$ is close to an integer n (n=1,2,3,4,5), with the criterion $|f-n| < 0.01 d^{-1}$; (2) the sum $(f_1+f_2)$ or difference $(|f_1-f_2|)$ of its first and second pulsation frequencies is close to an integer m (m=1,2,...,10). The tolerance window width for this frequency sum/difference is set to $0.005*m d^{-1}$ that widens with increasing m, i.e., if $||f_1\pm f_2|-m| < 0.005*m$, it is excluded.

Second, we screened on the amplitude of the secondary period $A_2$. After removing outliers with abnormally large amplitudes ($A_2>10$ mag), we required $A_2$ to exceed the photometric uncertainty $\sigma$ (typically $\sim$0.02 mag) for general candidates. For candidates with period ratios within the core sequence (0.742--0.748; the main RRd locus in Fig.~\ref{Fig.3}), recognizing that these sources are likely to be physical even when the secondary amplitude is small, we relaxed the criterion to $A_2>0.5\sigma$.

Finally, constraints on the fundamental pulsation period based on the physical properties of RRd stars are applied. Since RRd stars possess characteristics of both RRab (fundamental mode pulsation) and RRc (first-overtone pulsation), the typical fundamental pulsation period $P_{\rm F}$ of RRd stars should be in the range of 0.4 to 0.6 days. Therefore, we require the fundamental period of each candidate to satisfy $0.4 < P_{\rm F} < 0.6 $ d. This constraint effectively removes unreliable results where the fundamental period is close to 1 day (possibly affected by residual aliasing or non-physical) or shorter than 0.3 days (possibly corresponding to higher overtones or non-RRL sources).

\subsection{Searching for the Third Period}
\label{section3.3}
After successfully screening RRd stars, we further attempted to identify targets with three or more pulsation periods among them. Theoretically, if such variable stars exist, their period combinations should form additional sequences on the Petersen diagram. The search method is similar to the identification of the second period: for each double-mode candidate, we first use its two determined periods to perform a Fourier series fit, then obtain a secondary residual sequence through pre-whitening (i.e., subtracting the double-mode fitted model). Subsequently, LS analysis is applied to the secondary residual sequence to generate a new power spectrum to search for a potential third pulsation period $P_3$. During the search, we strictly excluded false signals related to the Earth's rotation period (1 day) or its integer multiples, as well as beat frequencies derived from $P_1$ and $P_2$, caused by the ZTF daily sampling window (screening criteria same as Section \ref{section3.2}). However, the analysis results showed that after effectively excluding these interferences, only a few individual candidates exhibited possible third period signals. These candidates are scarce and cannot be reliably distinguished by visual inspection; the physical authenticity of their third periods requires verification with higher precision and more densely sampled data. Referring to Fig.~\ref{Fig.2}, the amplitude of the second period is generally below 0.1 mag, and the third period amplitude is likely submerged in noise and indistinguishable. Therefore, we did not observe any new sequences on the Petersen diagram similar to those commonly seen in $\delta$ Scuti variable stars, formed by the combination of a third period with the first two \citep{2025ApJ...984...89J}.

\begin{table*}[t]
\centering
\caption{RRL star catalog of ZTF. }
\label{Table.1}
\setlength{\tabcolsep}{0pt} 
\begin{tabular*}{\textwidth}{@{\extracolsep{\fill}} c c c c c c c c c }
\hline
ID$_{GAIA}$ & RA & DEC & $P$ & Amp & $\log$(FAP) & ${\rm mag}_r$ & ${\rm mag_{err}}$ & $N_{obs}$ \\
-- & deg & deg & day & mag & -- & mag & mag & -- \\
\hline

177571127944832 & 46.05556 & 0.97448 & 0.6163544 & 0.160 & -99.1 & 18.169 & 0.040 & 777 \\
288243845193088 & 44.32286 & 0.78524 & 0.5545360 & 0.235 & -80.9 & 18.470 & 0.045 & 468 \\
500243431117184 & 44.69808 & 1.74565 & 0.6230946 & 0.272 & -70.0 & 18.161 & 0.038 & 418 \\
507222753405440 & 44.66533 & 1.78942 & 0.6088971 & 0.289 & -70.9 & 14.709 & 0.067 & 635 \\
584630948352256 & 46.34146 & 1.54204 & 0.5554842 & 0.313 & -69.6 & 17.462 & 0.024 & 421 \\
782027645388032 & 46.69908 & 2.30393 & 0.5583137 & 0.220 & -81.7 & 17.919 & 0.047 & 385 \\
1035533795140608 & 46.81719 & 3.02630 & 0.3519676 & 0.168 & -128.4 & 17.679 & 0.025 & 424 \\
1514169246023424 & 42.86076 & 2.45392 & 0.6143901 & 0.292 & -102.9 & 15.739 & 0.013 & 580 \\
1742622850845696 & 45.23365 & 2.93005 & 0.4564581 & 0.366 & -77.0 & 18.083 & 0.034 & 355 \\
2057461133775616 & 44.04663 & 3.64684 & 0.3488986 & 0.165 & -125.2 & 18.345 & 0.043 & 415 \\

\multicolumn{8}{c}{\vdots} \\
6915409547181662976 & 314.97175 & -3.83483 & 0.3466598 & 0.175 & -221.8 & 17.754 & 0.023 & 926 \\
6915461877061997312 & 316.04024 & -3.60833 & 0.5382781 & 0.272 & -77.1 & 18.005 & 0.032 & 395 \\
6915530184222277248 & 315.69290 & -3.42167 & 0.5203319 & 0.273 & -89.6 & 17.882 & 0.027 & 463 \\
6915656391836771712 & 313.97410 & -3.50264 & 0.4205020 & 0.155 & -74.0 & 17.258 & 0.020 & 466 \\
6915755584105531392 & 315.20562 & -3.00723 & 0.3814407 & 0.174 & -166.9 & 18.125 & 0.025 & 914 \\
6915794960365285376 & 314.51556 & -2.89618 & 0.6670621 & 0.283 & -97.2 & 17.492 & 0.065 & 622 \\
6915797777863835520 & 314.59083 & -2.89327 & 0.4827241 & 0.330 & -85.0 & 17.244 & 0.055 & 787 \\
6915849798508848384 & 316.60191 & -3.36223 & 0.6011085 & 0.190 & -95.2 & 18.166 & 0.031 & 473 \\
6915867498069057664 & 316.42845 & -3.19956 & 0.4897983 & 0.275 & -80.9 & 16.150 & 0.032 & 540 \\
6915881821785473792 & 316.67391 & -3.08883 & 0.6258177 & 0.201 & -90.5 & 17.566 & 0.022 & 485 \\

\hline

\end{tabular*}

\vspace{5pt}

\parbox{\linewidth}{\footnotesize \textbf{Note.} This table shows only a portion of the full dataset. The complete table is available in machine-readable format.}
\end{table*}
\section{results}
\label{section4}
Through the systematic screening pipeline, we ultimately constructed a catalog containing 38,524 RRL with reliable pulsation periods, amplitudes, magnitudes (including errors), FAP value and photometric point counts (Table \ref{Table.1}). This sample size is approximately 40\% of the total number of initially downloaded photometric data candidates. Among them, we successfully identified 969 RRd stars (Table \ref{Table.2}, comprising 918 RRd stars with $P_1$ corresponding to the first-overtone mode and 51 RRd stars with $P_1$ corresponding to the fundamental mode. Table \ref{Table.2} adopts the same column definitions as Table \ref{Table.1} but, for each RRd star, additionally reports the parameters of second period. Notably, the majority of RRd stars are newly discovered (626 stars), not included in the Gaia DR3 RRL catalog.

Fig.~\ref{Fig.2} shows the distribution of the entire RRL sample on the period-amplitude diagram. Gray points represent single-period RRL, which clearly separate into two main groups corresponding to RRab (longer periods and larger amplitudes) and RRc (shorter periods and smaller amplitudes) pulsators. We performed a simple classification based on the period--amplitude diagram. The resulting classifications differ from those in the Gaia catalog for only $\sim1\%$ of the sources, and these discrepancies almost exclusively occur near the boundaries of the RRab and RRc regions. Owing to the strict daily alias exclusion applied during the single-period star selection (see Section \ref{section3.2}), frequency regions close to integer multiples of one day ($1 d^{-1}, 2 d^{-1}, 3 d^{-1}$, etc.) inherently exhibit gaps in the diagram, which are unrelated to the intrinsic properties of RRL. To address this, we adopted the Gaia-provided frequencies as priors for RRL stars located within these alias-affected regions and reran the screening process in their vicinity. This procedure successfully recovered 798 stars, effectively filling the sampling-induced gaps, and these stars were added to Table \ref{Table.1}. 

In Fig.~\ref{Fig.2}, for RRd stars, red symbols indicate cases where $P_1$ corresponds to the fundamental mode, while blue symbols indicate cases where $P_1$ corresponds to the first-overtone mode. Cross symbols mark the location of $P_1$ and its amplitude for each RRd star, whereas triangles mark $P_2$ and its amplitude. A similar recovery procedure was applied to RRd stars: we used the frequencies provided by the Gaia RRd catalog as priors and reran the screening process with a relaxed frequency range, successfully recovering 33 RRd stars. These recovered stars have also been included in Table \ref{Table.2}. This effort partially fills the gaps that would otherwise appear in both the period--amplitude and Petersen diagrams.

\begin{figure*}
\centering
\begin{minipage}[b]{0.45\textwidth}
\centering
\includegraphics[width=\textwidth]{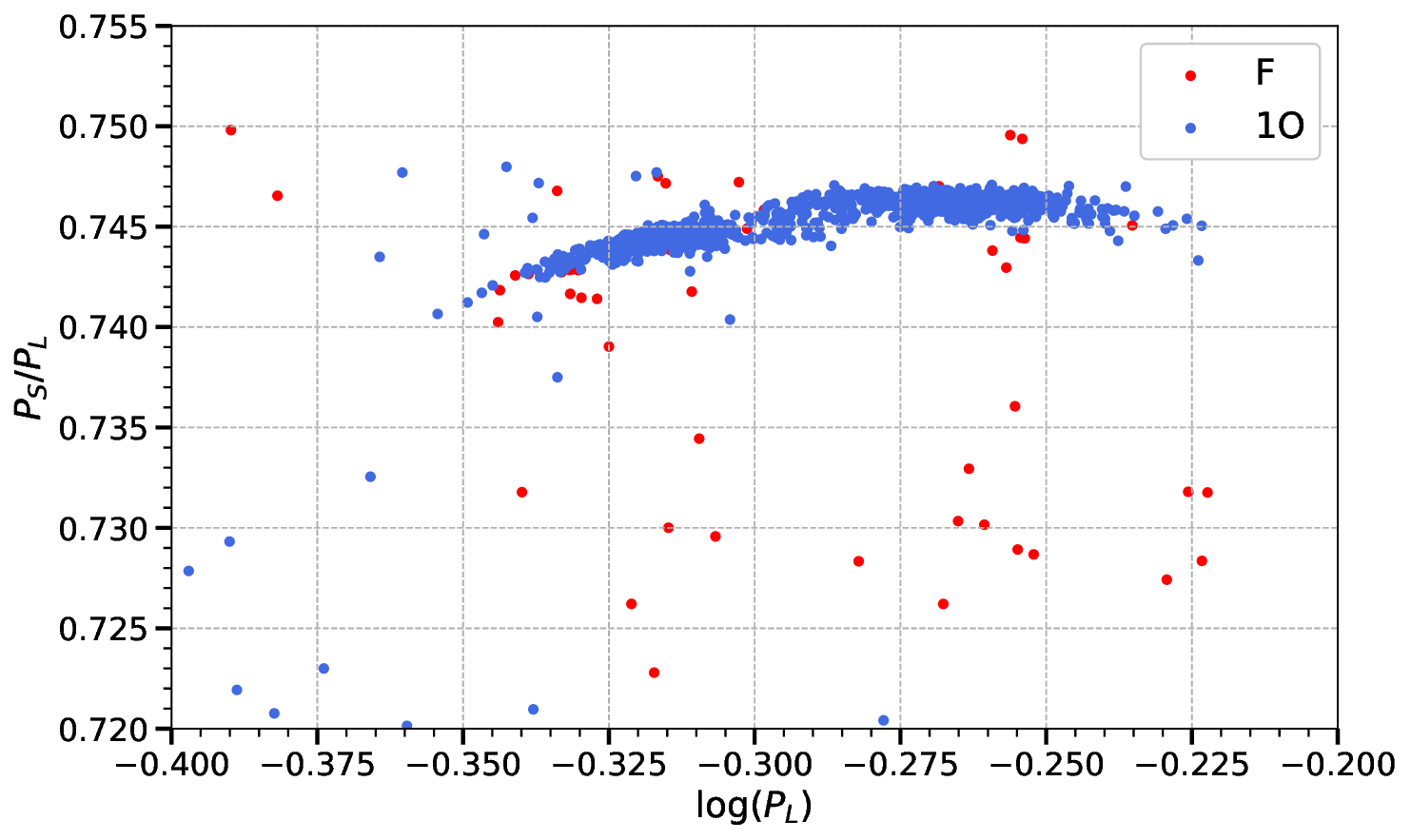}
\label{Fig.3_a}
\end{minipage}
\hfill
\begin{minipage}[b]{0.45\textwidth}
\centering
\includegraphics[width=\textwidth]{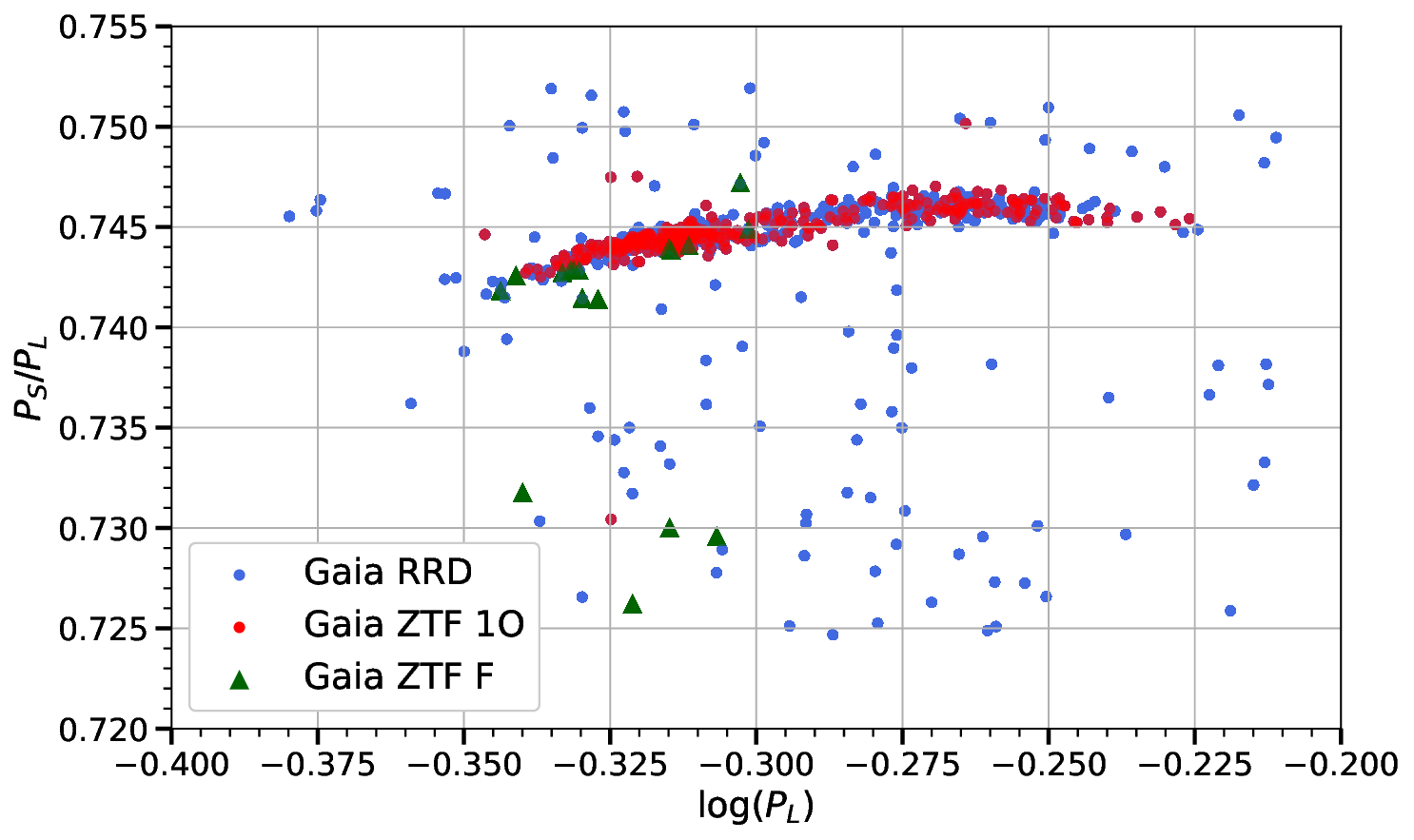}
\label{Fig.3_b}
\end{minipage}
\caption{Left Panel: Petersen diagram for RRd stars screened using ZTF data. Red points represent RRd stars where $P_1$ is the fundamental mode, blue points represent RRd stars where $P_1$ is the first-overtone mode. Blue points form a clear sequence while red points show no clear sequence and may be anomalous RRd stars. Right Panel: Petersen diagram for Gaia DR3 RRd stars with ZTF $r$-band coverage. Blue points: Gaia DR3 RRd subclass after cross-matching with ZTF. Red points: among the cross-matched set, RRd stars recovered by our pipeline where $P_1$ is the first-overtone mode. Green triangles: among the cross-matched set, RRd stars recovered by our pipeline where $P_1$ is the fundamental mode.}
\label{Fig.3}
\end{figure*}

Fig.~\ref{Fig.2} highlights several points. First, for RRd stars in both the red and blue groups, the $P_1$ points lie tightly within the locus occupied by single-mode RRL (gray), indicating that their dominant pulsation mode is consistent with that of their single-mode counterparts. Second, the secondary-period amplitude $A_2$ is systematically and substantially smaller than $A_1$, typically \(<0.1\) mag. Such weak amplitudes likely drive the signal-to-noise of additional (higher-overtone) modes below our detection threshold, providing a natural explanation for the absence of robust third-period detections in Section \ref{section3.3}. Notably, the $P_2$ points of the blue group (i.e., their fundamental-mode periods) exhibit a distinct clustering near $P\sim0.5$ d with amplitudes \(<0.1\)~mag. This forms a sparse, separately distributed locus relative to the gray single-mode RRL, reflecting markedly lower amplitudes than typical single-mode variables. The distinct distribution in the period--amplitude diagram can therefore serve as an auxiliary diagnostic for RRd screening, complementing the primary screening based on the period ratio sequence in the Petersen diagram (Fig.~\ref{Fig.3}). We also cross-matched our sample with the K2 RRd catalog from \citet{2024MNRAS.529..296N} and identified 20 overlapping RRd stars. These stars exhibit negligible period differences (on the order of $10^{-5}$ d) but show a systematic amplitude offset of 0.01 mag, with the K2 amplitudes being consistently higher. This offset is likely due to the fact that the K2 band is both broader and slightly bluer than the $r$ band, resulting in intrinsically larger observed amplitudes.

Fig.~\ref{Fig.3} presents the Petersen diagram, i.e. the period ratio $P_{1\mathrm{O}}/P_{\mathrm F}$ versus the fundamental period $P_{\mathrm F}$, for the screened RRd sample. RRd stars whose dominant mode is the first-overtone mode (blue) delineate a narrow, well-populated sequence and comprise the vast majority of the sample ($\sim$95\%). By contrast, systems dominated by the fundamental mode (red) are sparse and broadly dispersed without an obvious sequence, accounting for only $\sim$5\% of the sample. The scattered red points may represent non-typical RRd stars \citep{2023A&A...677A.177N,2016MNRAS.461.2934S}. Their period ratios may fall outside the typical interval (0.72-0.75) used in our screening. However, because ZTF’s $\sim$1-day cadence is prone to aliasing, we deliberately refrained from broadening this window; as a result, such stars are likely underrepresented in our sample.

\begin{table*}[t]
\centering
\caption{RRd star catalog of ZTF. }
\label{Table.2}
\setlength{\tabcolsep}{3pt} 
\begin{tabular*}{\textwidth}{@{\extracolsep{\fill}}  c c c c c c c c c c}
\hline
ID$_{GAIA}$ & RA & DEC & $P_1$ & $A_1$ & $P_2$ & $A_2$ & ${\rm mag}_r$ & ${\rm mag_{err}}$ & $N_{obs}$ \\

-- & deg & deg & day & mag & day & mag & mag & mag & -- \\
\hline

1035533795140608 & 46.81719 & 3.02630 & 0.3519676 & 0.168 & 0.4729896 & 0.040 & 17.679 & 0.025 & 424 \\
26659862734487040 & 40.21567 & 13.29430 & 0.3871893 & 0.175 & 0.5187844 & 0.030 & 16.407 & 0.014 & 1198 \\
27410550002747008 & 44.27695 & 10.91828 & 0.3525526 & 0.169 & 0.4739827 & 0.058 & 19.311 & 0.083 & 548 \\
27784319532128512 & 47.20757 & 12.07100 & 0.3523732 & 0.171 & 0.4817093 & 0.059 & 16.803 & 0.016 & 1010 \\
31122432539161216 & 48.35651 & 15.36304 & 0.3586350 & 0.173 & 0.4815118 & 0.048 & 17.208 & 0.026 & 2329 \\
36806327835701248 & 58.07698 & 11.87133 & 0.3493446 & 0.153 & 0.4697347 & 0.091 & 19.049 & 0.060 & 315 \\
42308726400739072 & 53.25898 & 15.19007 & 0.3485585 & 0.175 & 0.4688115 & 0.088 & 18.619 & 0.050 & 341 \\
51156844364167552 & 58.05123 & 20.35311 & 0.3638072 & 0.165 & 0.4887889 & 0.034 & 15.562 & 0.011 & 966 \\
54010936031517440 & 62.04137 & 24.12649 & 0.3934894 & 0.169 & 0.5270055 & 0.051 & 17.081 & 0.018 & 854 \\
59743824016820480 & 46.11214 & 18.60920 & 0.4094009 & 0.149 & 0.5485351 & 0.063 & 17.501 & 0.023 & 692 \\

\multicolumn{9}{c}{\vdots} \\

6898914506236972800 & 318.87825 & -6.38296 & 0.4083624 & 0.170 & 0.5469635 & 0.035 & 15.743 & 0.035 & 872 \\
6899434330423902080 & 320.84633 & -5.27805 & 0.4048619 & 0.171 & 0.5422459 & 0.043 & 18.429 & 0.036 & 472 \\
6899434330423902080 & 309.84121 & -11.55867 & 0.3635518 & 0.176 & 0.4879041 & 0.076 & 14.097 & 0.021 & 469 \\
6903740640729291264 & 311.58908 & -8.82224 & 0.4236938 & 0.154 & 0.5686013 & 0.081 & 15.515 & 0.041 & 536 \\
6904978312866242816 & 307.40200 & -9.36520 & 0.3447190 & 0.167 & 0.4641149 & 0.012 & 15.578 & 0.011 & 518 \\
6909445078852501888 & 315.36145 & -7.21891 & 0.3634751 & 0.161 & 0.4882881 & 0.124 & 17.073 & 0.017 & 444 \\
6913480252166900480 & 312.14227 & -5.49153 & 0.3638280 & 0.170 & 0.4883337 & 0.056 & 16.196 & 0.013 & 445 \\
6914919753406067328 & 313.37428 & -3.42978 & 0.4341056 & 0.154 & 0.5822662 & 0.050 & 14.068 & 0.025 & 482 \\
6915656391836771712 & 313.97410 & -3.50264 & 0.4205019 & 0.155 & 0.5640852 & 0.087 & 17.258 & 0.020 & 466 \\
6915755584105531392 & 315.20562 & -3.00722 & 0.3814407 & 0.174 & 0.5115608 & 0.025 & 18.125 & 0.025 & 914 \\

\hline
\end{tabular*}

\vspace{5pt}

\parbox{\linewidth}{\footnotesize \textbf{Note.} This table shows only a portion of the full dataset. The complete table is available in machine-readable format.}
\end{table*}

\section{discussion}
\label{section5}
\subsection{RRd region in the Period-Amplitude Diagram}
\label{section5.1}
To verify whether the independent clustered region mentioned in Section \ref{section4} on the period-amplitude diagram can serve as an auxiliary condition for RRd star screening, we analyzed whether single-period RRL exist in this region. The region was selected as $0.45 < P_{\rm F} < 0.6$ d and $A < 0.15$ mag, a condition that can cover all RRd stars within this range. We examined RRL within this region that were not identified as RRd stars. Excluding some RRc stars and those with poor photometric quality (few points or large errors), 35 RRL remained that exhibited long-period modulation of period and amplitude (i.e., the Blazhko effect \citep{1907AN....175..325B,1916ApJ....43..217S,2024AJ....167...14V,2024AJ....168...43V}). Due to the limited ZTF sampling cadence, our screening pipeline cannot robustly determine whether these objects are double-mode pulsators, leading to the omission of the corresponding candidates. These stars comprise less than 4\% of the candidates across the entire region and thus have only a minor impact on completeness. In summary, provided the data-quality filters are satisfied, this region can serve as an auxiliary criterion for confirming RRd star screening.

\begin{figure}
\centering
\includegraphics[width=0.45\textwidth]{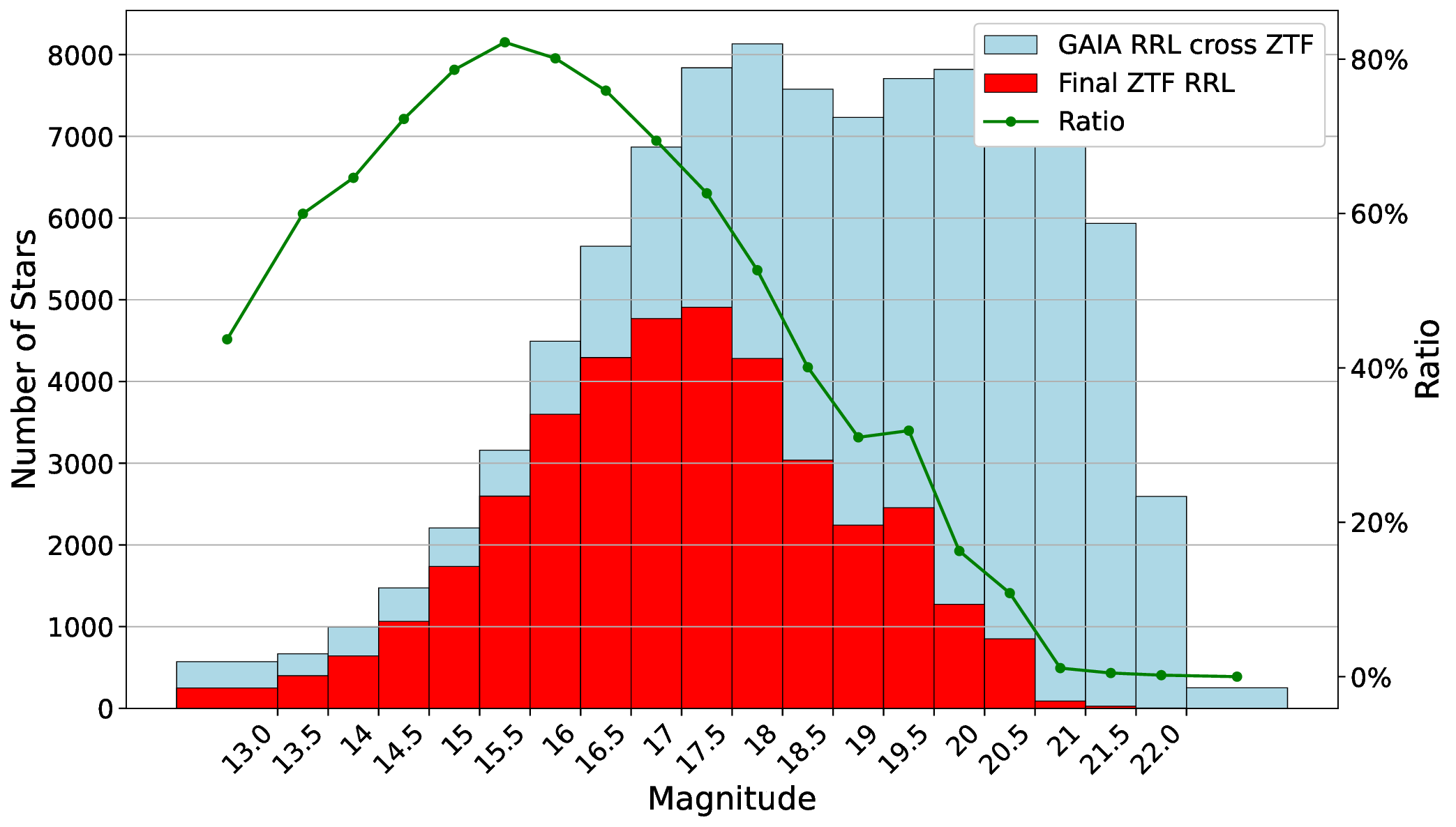}
\caption{\textbf{$r$-band magnitude histograms of the Gaia--ZTF cross-matched RRL catalog and the final RRL catalog from this work. The red histogram represents the final ZTF RRL catalog, while the blue one corresponds to the Gaia--ZTF cross-matched RRL sample. The bin size is 0.5 mag, with sources brighter than 13~mag and fainter than 22~mag grouped into single bins at each end. The green line indicates the completeness, defined as the ratio of the red to blue counts in each bin. This figure demonstrates how the catalog completeness varies as a function of magnitude.}}
\label{Fig.4}
\end{figure}

\subsection{Completeness of the RRL Catalog}
\label{section5.2}
The RRL catalog obtained in Section~\ref{section4} contains only 40\% of the cross-matched sample between ZTF DR22 and Gaia DR3. To investigate the cause of this incompleteness, we constructed magnitude histograms with 0.5~mag bins for both the full set of 97,146 cross-matched candidate stars and the final catalog of 39,322 RRL stars, and computed the completeness in each magnitude bin (i.e., the ratio of the number of cataloged RRL stars to the number of cross-matched candidates; see Fig.~\ref{Fig.4}). The completeness peaks at 15--15.5~mag, reaching approximately 80\%, but declines toward both brighter and fainter magnitudes, dropping to nearly zero beyond 20~mag. This strong dependence on magnitude arises primarily because ZTF photometric precision is lower than Gaia’s for fainter stars, making variability harder to detect, while stars brighter than $\sim$12.5~mag are saturated in ZTF images, which also leads to missed detections. In addition to these magnitude-related effects, other factors—including spatial inhomogeneities in ZTF photometry, the sparse temporal sampling intrinsic to wide-field surveys, and ZTF’s $\sim$1-day cadence (which is close to the typical RRL period range of 0.2--0.8~days)—account for the remaining $\sim$20\% of incompleteness. A more detailed investigation of the completeness of the ZTF RRL sample is beyond the scope of this work, which instead focuses on improving the completeness of RRd star identification.

\subsection{Completeness of the RRd Catalog}
\label{section5.3}
To quantify the completeness of our RRd star screening, we performed a cross-validation against Gaia DR3 (Fig.~\ref{Fig.3}). From the Gaia DR3 RRL catalog, we selected sources classified as RRd stars and having ZTF DR22 $r$-band photometric data, yielding 745 cross-matched objects (blue points). Our pipeline ultimately recovered 355 of these as RRd stars (red points for cases with $P_1$ as the first-overtone mode; black points for cases with $P_1$ as the fundamental mode), corresponding to a completeness of 47.7\%, which exceeds the 40\% completeness achieved for the full RRL sample. For the recovered sources, our period measurements closely match the Gaia DR3 values, with a typical (median) absolute difference $<0.001$ d, validating the accuracy of our catalog periods.

An in-depth analysis of the Gaia-labeled RRd stars that were not identified by our pipeline reveals two main causes. First, 327 objects were not included in our ZTF-based parent catalog of 39,322 RRL because no significant pulsation period could be determined, primarily due to sparse ZTF sampling or relatively large photometric uncertainties. For the remaining 63 Gaia RRd stars that were not identified by our method, we examined their ZTF light curve using the second period provided by Gaia. Among them, three stars have identical Gaia fundamental and first-overtone periods, indicating that they were misclassified as RRd stars in Gaia. Eighteen stars exhibit excessively high noise levels in the power spectra around Gaia’s second frequency, mainly due to an insufficient number of valid ZTF observations. For the remaining 42 stars, even when the noise level near Gaia’s second frequency is not particularly high, no significant power peaks are detected. We suspect that the Gaia second periods for these stars are likely erroneous, given that Gaia typically provides only about 40 photometric measurements per source; however, we cannot rule out the possibility that the ZTF photometry is still too sparse to detect the secondary signal. For these 42 stars, we also consulted the ASAS-SN classifications, which do not show evidence of double-mode pulsations. We list the 45 RRd stars with potentially incorrect Gaia second periods in Table~\ref{Table.3} to facilitate future improvements. The table includes Gaia’s fundamental and first-overtone periods, the ZTF-measured periods, and the ASAS-SN periods and classifications \citep{2018MNRAS.477.3145J,2019MNRAS.486.1907J,2021MNRAS.503..200J}

\begin{table*}[t]
\centering
\caption{Catalog of stars in the Gaia RRd sample with potentially spurious secondary periods}
\label{Table.3}
\setlength{\tabcolsep}{3pt} 
\begin{tabular*}{\textwidth}{@{\extracolsep{\fill}}  c c c c c c c c c c}
\hline
\multicolumn{1}{c}{ID$_{\rm Gaia}$} & \multicolumn{1}{c}{RA}  & \multicolumn{1}{c}{DEC} & \multicolumn{1}{c}{$P_{\rm ZTF}$} & \multicolumn{1}{c}{$P_{\rm F\_{Gaia}}$} & \multicolumn{1}{c}{$P_{\rm 1O\_{ Gaia}}$} & \multicolumn{1}{c}{$mag_r$} & \multicolumn{1}{c}{$P_{\rm ASASSN}$} & \multicolumn{1}{c}{$\rm Type_{\rm ASASSN}$} \\
\multicolumn{1}{c}{--}       & \multicolumn{1}{c}{deg} & \multicolumn{1}{c}{deg} & \multicolumn{1}{c}{day}    & \multicolumn{1}{c}{day}        & \multicolumn{1}{c}{day}         & \multicolumn{1}{c}{mag} & \multicolumn{1}{c}{day}       & \multicolumn{1}{c}{--}           \\
\hline
502860822626623104  & 85.72881  & 74.78611  & 0.6041705 & 0.6041509           & 0.4385321           & 17.368 &           &                \\
1450617851240322688 & 208.40968 & 26.35033  & 0.5797049 & 0.5797297           & 0.4230145           & 16.346 & 0.5796831 & RRAB           \\
1596792738128894592 & 236.76926 & 53.1561   & 0.4466364 & 0.6096576           & 0.4463491           & 14.394 & 0.4466521 & RRC            \\
1783920408766381952 & 320.39789 & 16.13584  & 0.4602743 & 0.4602788           & 0.3361554           & 15.63  & 0.4602751 & RRAB           \\
1805848274101335040 & 307.03949 & 16.27775  & 0.4913785 & 0.4914004           & 0.362823            & 17.677 &           &                \\
1810730037011464960 & 308.6307  & 15.17989  & 0.4915126 & 0.4914803           & 0.361803            & 17.857 &           &                \\
2031904123879963520 & 297.0744  & 29.85339  & 0.421512  & 0.4215087           & 0.3143631           & 20.215 &           &                \\
2090026915546876416 & 280.1191  & 31.19779  & 0.5222852 & 0.5222908           & 0.3844919           & 17.134 &           &                \\
2115356055676066688 & 273.62917 & 44.6829   & 0.5505965 & 0.550598            & 0.4004461           & 17.193 & 0.5505671 & RRAB           \\
2758708243554423040 & 357.95006 & 8.5922    & 0.5285533 & 0.5285581           & 0.3930872           & 18.418 &           &                \\
2795019722436672000 & 8.45958   & 19.09664  & 0.4693872 & 0.4693932           & 0.3454607           & 16.101 &           &                \\
2814405448168815104 & 351.94422 & 15.64519  & 0.3577714 & 0.4828938           & 0.3577731           & 19.894 &           &                \\
3083608745540611712 & 122.04698 & 0.04515   & 0.612231  & 0.6122232           & 0.4580622           & 16.671 & 0.5246499 & RRD            \\
3213529509080988928 & 79.09212  & -3.85864  & 0.3639642 & 0.4887554           & 0.3639641           & 18.605 &           &                \\
3309315457399564160 & 68.33308  & 14.18479  & 0.5428323 & 0.5428196           & 0.4044056           & 19.761 &           &                \\
3517694458714925056 & 181.6626  & -20.58981 & 0.5290777 & 0.5290455           & 0.3941454           & 17.912 &           &             \\
3608077643117446784 & 200.9125  & -14.70316 & 0.5257874 & 0.5257962           & 0.3813279           & 15.948 & 0.5258136 & RRAB           \\
3682664874108530944 & 191.70032 & -2.12195  & 0.6123337 & 0.6123072           & 0.4489812           & 15.439 & 0.6123273 & RRAB           \\
3709729589048951808 & 192.71755 & 7.6519    & 0.5314695 & 0.5314589           & 0.3884121           & 18.977 &           &                \\
3795393673840361728 & 178.49371 & 0.87337   & 0.3833001 & 0.5140027           & 0.3832835           & 19.708 &           &                \\
3852201698731357440 & 141.5509  & 5.21065   & 0.5296877 & 0.5296718           & 0.3862219           & 14.946 & 0.5296829 & RRAB           \\
3922379123590279168 & 181.90177 & 15.46989  & 0.3579839 & 0.4809577           & 0.35797             & 19.13  &           &                \\
4196129554383078272 & 299.79252 & -6.69481  & 0.4421725 & 0.442187            & 0.3301647           & 15.409 & 0.44217   & RRAB           \\
4196457513785766784 & 297.25045 & -7.05786  & 0.520707  & 0.5207136           & 0.3894893           & 19.599 &           &                \\
4240422727473432320 & 297.50637 & 0.68016   & 0.5291321 & 0.5291366           & 0.3910091           & 17.669 &           &                \\
4330255022182722688 & 248.02916 & -14.00887 & 0.5290188 & 0.5290096           & 0.3951421           & 19.752 &           &                \\
4353777084891331200 & 252.04701 & -4.10632  & 0.4815344 & 0.4815563           & 0.3597401           & 18.144 &           &                \\
4354942842094568576 & 249.82499 & -2.81677  & 0.5252633 & 0.5252557           & 0.3822973           & 17.304 &           &             \\
4378643227547650560 & 251.55756 & -3.35357  & 0.5297852 & 0.529778            & 0.3918242           & 16.849 &           &                \\
4396426518159609088 & 237.37217 & -6.96432  & 0.4113024 & 0.5511965           & 0.4112925           & 19.186 &           &                \\
4444659825527857024 & 256.76491 & 10.63418  & 0.4890972 & 0.489091            & 0.3668659           & 17.595 &           &             \\
4473567261831119744 & 264.8382  & 4.82405   & 0.5328462 & 0.5328471           & 0.3932263           & 16.3   & 0.5328481 & RRAB           \\
4488253850960909952 & 266.5316  & 8.82408   & 0.4434247 & 0.4434342           & 0.3310913           & 15.09  & 0.4434243 & RRAB           \\
4550557402469209344 & 264.61843 & 18.25681  & 0.5490264 & 0.5490168           & 0.3979703           & 14.22  & 0.5490296 & RRAB           \\
4605919698416163584 & 274.2926  & 35.90056  & 0.5480294 & 0.5480033           & 0.3997937           & 17.751 & 0.5480327 & RRAB           \\
5456821924563199872 & 165.29836 & -27.70694 & 0.3842179 & 0.5154945           & 0.3842201           & 19.387 &           &                \\
5651867071056411520 & 136.91575 & -22.90902 & 0.3675377 & 0.4932827           & 0.367545            & 17.953 &           &             \\
5702666226247730688 & 130.67119 & -21.12655 & 0.613267  & 0.6132665           & 0.4520622           & 18.91  &           &                \\
5712267402298562560 & 118.25115 & -21.68047 & 0.3725988 & 0.5005112           & 0.3726196           & 19.4   &           &                \\
5717043886969335296 & 114.50211 & -18.69806 & 0.4768092 & 0.476815            & 0.3504548           & 19.927 &           &                \\
6248369549512371712 & 240.34383 & -18.23853 & 0.5623893 & 0.5623407           & 0.4222875           & 19.718 &           &                \\
6808559866186937216 & 313.83632 & -21.87428 & 0.5297517 & 0.5297671           & 0.3930037           & 14.632 & 0.5297643 & RRAB           \\
6863908117597473408 & 298.11201 & -23.29828 & 0.4544423 & 0.4539646           & 0.3366025           & 15.247 & 0.3366015 & RRC            \\
3067012575294029824 & 121.12661 & -6.46421  & 0.406481  & 0.4064984 & 0.4065198 & 19.573 &           &                \\
4512152178571784576 & 281.38468 & 17.71024  & 0.4085317 & 0.4085248  & 0.4085263 & 17.179 &           &                \\
6885270662153363328 & 315.5748  & -14.3219  & 0.4083273 & 0.4083270  & 0.4083259 & 19.918 &           &               \\
\hline
\end{tabular*}

\vspace{5pt}

\end{table*}

Furthermore, Fig.~\ref{Fig.3} shows an interesting phenomenon: near the frequency of approximately 2 $d^{-1}$ (corresponding to a period of 0.5 days), the density of the Gaia RRd stars (blue points) also shows a decrease. Given that the Gaia space telescope is located at the L2 point, its observations are not affected by Earth's rotation period, so this sparse region is unlikely to be caused by daily sampling periods and more likely reflects the intrinsic metallicity distribution characteristics of RRd stars in ZTF survey, indicating a relatively scarce number of stars in this specific pulsation frequency region.

\begin{figure}
\centering
\includegraphics[width=0.45\textwidth]{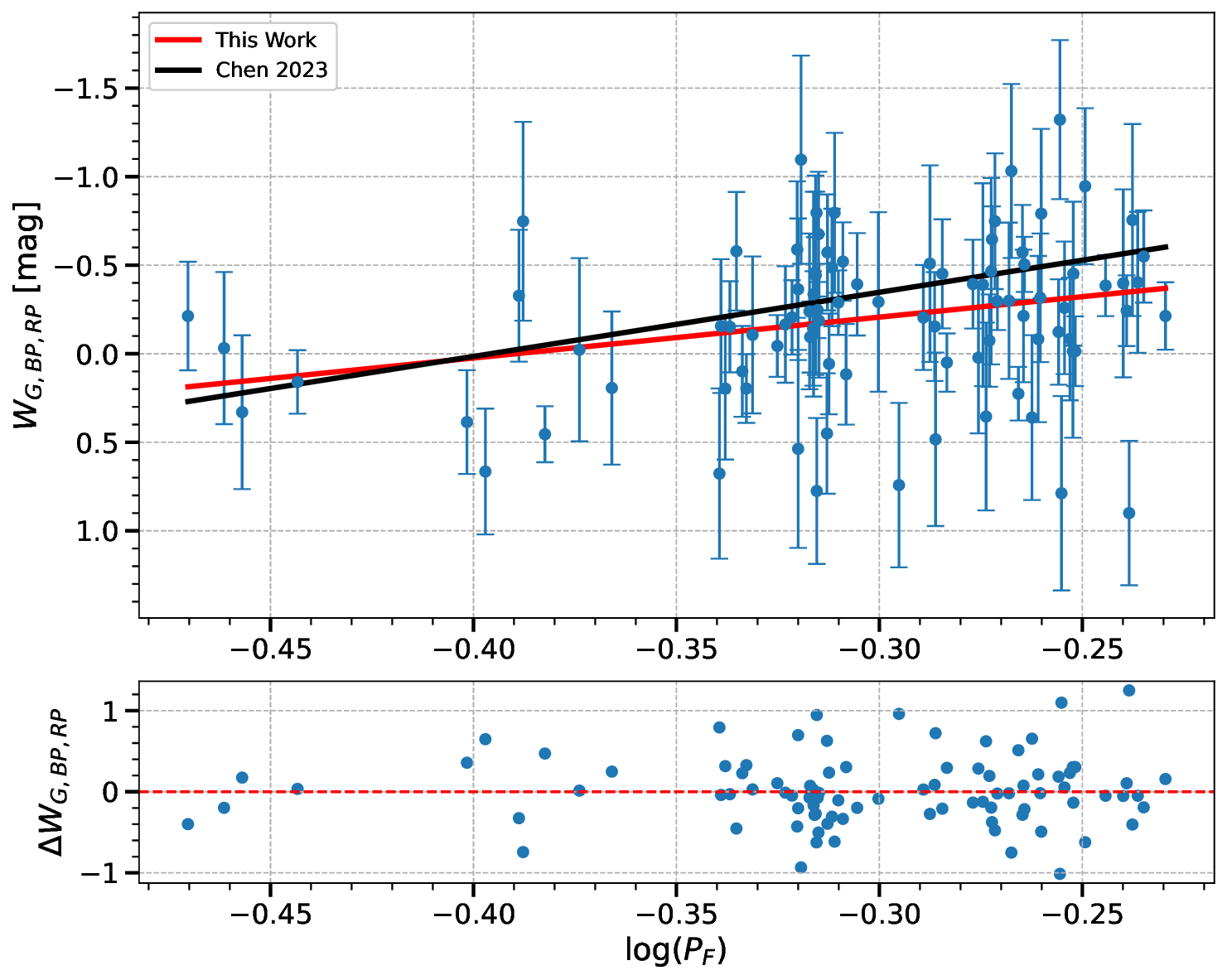}
\caption{Period-luminosity relation diagram for 90 RRd stars. This diagram is based on the Wesenheit magnitudes from Gaia's G, BP, and RP bands plotted against the fundamental-mode period. The red solid line represents the fitting result of this work, and the black solid line represents the result of \citet{2023NatAs...7.1081C}, where the distance modulus of the Large Magellanic Cloud is taken as 18.476 mag \citep{2019Natur.567..200P}. The lower panel shows a scatter plot of the residuals from the PLR.}
\label{Fig.5}
\end{figure}

\subsection{Period-Luminosity relation of RRd}
\label{section5.4}

To determine the PLR based on the RRd stars found in this work, we used Gaia parallaxes together with Wesenheit magnitudes constructed from its three bands, defined as $W_{G, BP, RP}= G -1.90(BP - RP)$ \citep{2019A&A...625A..14R, 2019ApJ...877..116W}. The parallaxes were corrected for the zero-point offset following the prescription of \citet{2021A&A...649A...2L}. We excluded stars with parallaxes smaller than five times their uncertainties, and removed $3\sigma$ outliers. After these quality cuts, a final sample of 90 RRd stars was retained, from which we derived the following PLR:

\[
\begin{aligned}
M_{W_{G, BP, RP}} &= (-2.310 \pm 0.584)(\log P_{\mathrm{F}}+0.35) \\
&\quad + (-0.091 \pm 0.040) \quad \sigma = 0.428 \, \mathrm{mag}
\end{aligned}
\]

Fig.~\ref{Fig.5} presents the PLR diagram, where the error bars account for uncertainties from both the photometric measurements and the parallaxes. The solid black line shows the PLR for RRd stars in \citet{2023NatAs...7.1081C}, and the red line is for the results of this work.
At the representative period of $\log P_{\rm F} = -0.35$, the difference between the two relations is only $0.08 \pm 0.11$~mag, indicating good consistency in the zero point. The zero point derived in this work is 0.08~mag larger than that of the Large Magellanic Clouds (LMC), primarily due to the Gaia parallax zero-point offset. \citet{2021ApJ...911L..20R,2022ApJ...938...36R,2023NatAs...7.1081C} found that even after applying the zero-point correction, Gaia parallaxes still exhibit residual systematics of 4--10~$\mu$as. Assuming that the RRd PLR in the LMC is accurate and adopting an average parallax of 0.2~mas for the 90 RRd stars in our sample, a zero-point difference of 0.08~mag corresponds to an 8~$\mu$as parallax offset, which is fully consistent with previous studies.

The standard deviation of our PLR is $\sigma = 0.428$~mag, which is relatively large and mainly driven by parallax uncertainties. Since the RRd stars identified in this work are, on average, located at larger distances, our constraints on the PLR slope and zero point are weaker than those from the LMC sample. In the future, by combining the RRd stars identified here with those discovered by Gaia and other surveys, as well as with the LMC RRd population, and by applying further refined Gaia parallax corrections, the zero point of the RRd PLR can be more tightly constrained. This improved calibration will enable more accurate distance measurements to dwarf galaxies.

\section{Conclusions}
\label{section6}
\par
By cross-matching the Gaia DR3 RRL catalog with ZTF DR22 time-series photometry, we assembled a catalog of 39322 RRL with reliably measured pulsation periods, amplitudes, mean magnitudes, and photometric information. In this sample, we identified 969 RRd stars, corresponding to 2.5\% of all RRL. This fraction is comparable to the $\sim$3\% toward the Galactic anti-center, higher than the $\sim$1\% in the Galactic bulge, and lower than the 5--10\% observed in the LMC and SMC \citep{2019AcA....69..321S}. More than half of these RRd stars are newly recognized and were not flagged as double-mode in the Gaia DR3 catalog, thereby validating the efficacy of our automated RRd-screening pipeline. For each RRd star, we provide the two pulsation periods, their associated amplitudes, FAP values, and magnitudes. Extrapolating from the recovered RRd fraction, we estimate that Gaia DR4 could yield at least $\sim$6,700 RRd stars, provided sufficient time-sampling for the RRL candidates in DR3 catalog.

Analysis of the period-amplitude diagram revealed a potentially important feature: for the first-overtone-dominated RRd stars (i.e., with $P_1$ as the dominant mode), their second periods and amplitudes occupy a compact, relatively isolated locus with respect to the global distribution of single-mode RRL. This clustering can serve as a useful auxiliary diagnostic for the RRd star screening. In the Petersen diagram, the dominant sequence is formed by first-overtone-dominated RRd stars, whereas fundamental-mode-dominated RRd stars are more sparsely distributed and do not form a clear sequence. RRd stars deviating from the sequence may belong to non-typical or anomalous RRd stars. Furthermore, the search for the third period among the identified RRd stars did not reveal any reliable pulsation signals, consistent with the generally weak amplitude of the second period.

To validate the completeness of our RRd star screening, we performed a cross-match with the Gaia DR3 RRd catalog. Among the 745 cross-matched RRd stars, our method successfully recovered 355 stars, corresponding to a recovery rate of 47.7\%. Of the remaining sources, 327 stars were not included in our parent RRL catalog. Further analysis of the other 63 stars revealed that 18 stars could not be correctly identified due to an insufficient number of valid ZTF observations, while the remaining 45 stars are potentially attributed to incorrect measurements of the second period in Gaia. Despite these limitations, we uncovered a substantial number of new RRd stars, underscoring the strong potential of time-domain surveys to reveal a larger true fraction of double-mode pulsators than commonly assumed.

Based on 90 RRd stars, we derived a PLR that is consistent with the LMC-based relation, showing only a small zero-point offset of $0.08\pm 0.11$ mag. This offset is likely caused by a zero-point bias of approximately $\sim$ 8 $\mu$as in Gaia parallaxes. Owing to the large distances of the sample, the scatter of the PLR ($\sigma = 0.428$ mag) is primarily driven by parallax uncertainties.

The enlarged RRd sample—especially given the weak metallicity dependence of their PLR—provides valuable leverage for precision distance determinations, studies of the Milky Way’s structure, and constraints on stellar evolution. Looking ahead, wide-field time-domain surveys from space and the ground (e.g., Legacy Survey of Space and Time \citep[LSST;][]{2019ApJ...873..111I}, China Space Station Telescope \citep[CSST;][]{2025arXiv250704618C}) will enable more complete and homogeneous censuses of RRd stars in the Local Group, thereby enhancing their utility for cosmological distance measurements.

\section*{Acknowledgements}
We thank the anonymous referee for the helpful comments. This work was supported by the National Natural Science Foundation of China (NSFC) through grants 12173047, 12322306, 12373028, 12233009 and 12133002. We also thanked the support from the National Key Research and development Program of China, grants 2022YFF0503404. X. Chen and S. Wang acknowledge support from the Youth Innovation Promotion Association of the Chinese Academy of Sciences (CAS, No. 2022055 and 2023065). This work is based on observations obtained with the 48 inch Samuel Oschin Telescope and the 60 inch Telescope at the Palomar Observatory as part of the Zwicky Transient Facility project. ZTF is supported by the National Science Foundation under grant Nos. AST-1440341 and AST-2034437 and a collaboration including current partners Caltech, IPAC, the Weizmann Institute for Science, the Oskar Klein Center at Stockholm University, the University of Maryland, Deutsches Elektronen-Synchrotron and Hum- boldt University, the TANGO Consortium of Taiwan, the University of Wisconsin at Milwaukee, Trinity College Dublin, Lawrence Livermore National Laboratories, IN2P3, University of Warwick, Ruhr University Bochum, Northwestern University and former partners the University of Washington, Los Alamos National Laboratories, and Lawrence Berkeley National Laboratories. Operations are conducted by COO, IPAC, and UW.

SOFTWARE: Astropy \citep{2013A&A...558A..33A,2018AJ....156..123A,2022ApJ...935..167A}, Matplotlib \citep{2007CSE.....9...90H}, ztfquery \citep{mickael_rigault_2018_1345222}
\renewcommand{\bibname}{References}
\bibliographystyle{aasjournal}
\bibliography{RRd}
\end{document}